# Macroeconomic Forecasting from Input-Output Tables Alone: A Darwinian Agent-Based Approach with FIGARO Data

Martin Jaraiz

Department of Electronics, University of Valladolid, Spain

## *Abstract*

How much macroeconomic information is contained in a single input-output table? We address this question by feeding FIGARO 64-sector symmetric industry-by-industry tables into DEPLOYERS, a Darwinian agent-based simulator, and using the result to produce genuine out-of-sample GDP forecasts. For each year, the model reads one FIGARO table for year *N*, self-organizes an artificial economy through evolutionary natural selection, then runs 12 months of fully autonomous free-market dynamics whose emergent growth rate predicts year *N*+1. The I-O table is essentially the **only** input: no time series, no estimated parameters, no expectations formation, no external forecasts.

We present five sets of results. First, a **9-year Austrian panel (2010–2018)** using 12-seed ensembles produces MAE of 1.22 pp overall; for five non-crisis years, MAE falls to **0.42 pp**—comparable to the best professional forecaster (WIFO: 0.48 pp). A **Swedish 9-year panel** independently confirms this accuracy (normal-years MAE 0.80 pp). Second, **cross-country portability** is demonstrated: 33 of 37 tested FIGARO countries converge with zero parameter changes. Third, a **German 9-year panel** reveals a systematic +3.7 pp positive bias from export dependency, and a **French panel** shows similar bias (+2.76 pp)—informative negative results pointing to multi-country network simulation as the natural extension. Fourth, a **COVID-19 simulation** demonstrates the I-O structure as a shock propagation mechanism: a 19-month timeline produces Year 1 GDP −4.62% versus empirical −6.6%. Fifth, **emergent firm size distributions** match European Commission data without micro-target calibration.

These results establish the I-O table as serving a dual purpose: structural baseline engine *and* dynamic shock propagation mechanism. The approach works best for economies with moderate growth where the structural trajectory closely tracks actual outcomes (Austria, Sweden); for economies with persistently low growth or high export dependency (France, Germany), a household wealth-consumption feedback mechanism produces a positive bias that represents a research frontier. Since FIGARO covers 46 countries with identical methodology, the approach is immediately portable without retuning parameters.

## 1. Introduction

Macroeconomic forecasting remains one of the most challenging tasks in applied economics. Despite decades of methodological development—from structural equation models through vector autoregressions to dynamic stochastic general equilibrium (DSGE) frameworks—forecast accuracy has improved only modestly. The International Monetary Fund's one-year-ahead GDP growth forecasts for advanced economies exhibit a mean absolute error of approximately 1.3 percentage points (Ismail, Perrelli and Yang, 2021). Consensus Economics surveys, aggregating dozens of professional forecasters, achieve roughly 1.0 pp (Federal Reserve Bank of St. Louis, 2024). The OECD's Economic Outlook forecasts show similar accuracy at approximately 1.5 pp (Pain et al., 2014). DSGE models, the workhorse of central bank forecasting, have been shown to be no more accurate than simple reduced-form benchmarks at horizons beyond one quarter (Edge and Gürkaynak, 2010; Smets and Wouters, 2007).

Against this backdrop, agent-based models (ABMs) have emerged as a promising alternative paradigm. Farmer and Foley (2009) argued in *Nature* that "the economy needs agent-based modelling" precisely because ABMs can capture the heterogeneity, nonlinear interactions, and out-of-equilibrium dynamics that representative-agent models suppress. The recent 90-page landmark survey by Axtell and Farmer (2025) in the *Journal of Economic Literature* documents the maturation of ABM methodology from a niche computational curiosity into a serious quantitative tool. A watershed moment came with Poledna, Miess, Hommes and Rabitsch (2023), who demonstrated in the *European Economic Review* that an ABM could pass the institutional benchmark test—producing GDP forecasts for Austria competitive with the IMF, OECD, and domestic forecasters. This result has catalyzed rapid developments: the Bank of Canada's CANVAS model (Hamill, Wieland and Hommes, 2025), the Bank of Italy's BeforeIT.jl framework (Glielmo, Devetak, Meligrana and Poledna, 2025), and the crisis-forecasting extension by Hommes and Poledna (2026) all build on the proof of concept that ABMs can be quantitative forecasting instruments.

This paper takes a different approach to the same question. Where Poledna et al. (2023) built a highly detailed model with 8.8 million agents, pre-calibrated from census microdata and business demography registers, requiring a supercomputer cluster and 500 Monte Carlo simulations, we ask: *how simple can an ABM be and still produce competitive forecasts?* Our answer is radically simple. DEPLOYERS uses approximately 2,000 agents on a standard desktop PC, reads a single FIGARO input-output table as its only structural input, estimates zero parameters, and lets Darwinian natural selection—firm birth when unmet demand is observed, firm death when wealth turns negative—organize the artificial economy. A complete 12-seed ensemble for one country-year runs overnight on a laptop.

The simplicity is deliberate and methodologically motivated. By stripping the model to its bare essentials—Leontief production technology, bilateral price tâtonnement, evolutionary firm dynamics, and an I-O table—we can isolate the contribution of the I-O data itself. If the resulting forecasts are competitive, the I-O table must contain more macroeconomic information than is conventionally extracted through standard Leontief demand-pull or Ghosh supply-push multiplier analysis.

We advance four claims:

1. **I-O tables are informationally rich for structural forecasting.** A single FIGARO 64-sector I-O table contains enough structural information—inter-sectoral technology, factor income distribution, household demand, government accounts, trade patterns—to generate one-year-ahead GDP forecasts comparable to professional institutions for years where no major exogenous shocks intervene.
2. **Darwinian selection is a sufficient equilibrating mechanism.** The evolutionary process of firm birth and death, operating on the I-O structural constraints, drives the economy to a steady state consistent with the table and then produces autonomous forward dynamics. No expectations formation, no Taylor rule, no estimated behavioral parameters are needed.
3. **Zero-parameter multi-country portability is achievable.** The same code, with the same universal behavioral rules, produces meaningful forecasts across structurally diverse economies by changing only the I-O table input. Of 37 tested FIGARO countries, 33 converge with zero parameter changes; full 9-year panels for Austria, Sweden, Germany, and France confirm the approach across different economic structures.
4. **The I-O network serves as a shock propagation mechanism.** The same inter-sectoral linkages that drive baseline forecast accuracy also faithfully transmit encoded exogenous shocks—as demonstrated with a COVID-19 pandemic simulation for Austria.

Importantly, we maintain honest framing throughout. DEPLOYERS predicts *structural trajectories*—what the economy would do based on its production structure absent in-year perturbations. This is comparable to what institutions predict only in years where the structural trajectory coincides with the actual outcome. On those years, our accuracy is comparable to the best professional forecasters; we do not claim to outperform them in general.

The paper proceeds as follows. Section 2 describes the model architecture and its relationship to the I-O table. Section 3 details the out-of-sample forecasting protocol. Section 4 presents five sets of results: the Austrian 9-year panel, the institutional comparison, cross-country portability, the German 9-year panel, the COVID-19 shock simulation, and emergent microeconomic structure. Section 5 discusses what the results reveal about I-O information content, the comparison with Poledna et al. (2023), the German puzzle, simplicity as robustness, and limitations. Section 6 concludes.

## 2. The DEPLOYERS Model

### 2.1 Darwinian Deployment Philosophy

DEPLOYERS belongs to the evolutionary economics tradition inaugurated by Nelson and Winter (1982), which treats economic change as an evolutionary process of variation, selection, and retention rather than an optimization problem solved by representative agents with rational expectations. The bounded rationality framework of Simon (1956)—agents satisfice rather than optimize—provides the behavioral foundation: firms and workers follow simple heuristic rules, and the economy's macroeconomic properties emerge from the interaction of these boundedly rational agents through market mechanisms.

The name "DEPLOYERS" derives from Jaraiz (2020), who developed a general framework for *self-organizing agent deployers* applicable to systems as diverse as ant colonies, robot coordination in industrial assembly, and economic production. The key insight is that complex systems that must allocate heterogeneous agents to heterogeneous tasks share a common organizational problem: how to achieve efficient aggregate outcomes without centralized planning. In ant colonies, pheromone trails provide the coordination signal; in robot assembly, task completion signals drive reallocation; in economies, prices and profits serve the same role. The same simulation code, with different input files describing the environment, can coordinate robots doing parallel assembly tasks or firms producing goods in a 64-sector economy. This domain generality is not an engineering curiosity but a conceptual point: the coordination mechanism is independent of the domain, and the structural input (task description for robots, I-O table for economies) determines the emergent behavior.

The economy-as-ecosystem metaphor has a specific technical implication. Unlike Poledna et al. (2023), who pre-calibrate the firm population, employment distribution, and wealth profiles from microdata before running the model, DEPLOYERS starts from the I-O table and lets the population *emerge*. Firms are born when entrepreneurs observe unmet demand (variation), compete for workers, inputs, and customers (selection), and survive or perish based on accumulated wealth (retention). The steady-state economy that results is one where the number of firms per sector, their sizes, and the employment distribution are all endogenously determined by the I-O structural constraints—not externally imposed. This is a genuinely novel feature in the ABM forecasting literature: no other macroeconomic ABM that we are aware of achieves competitive GDP forecasts with fully endogenous firm dynamics.

## 2.2 Model Structure and Data

Each forecast exercise begins with a single FIGARO 25th-edition symmetric industry-by-industry table. The 64-sector granularity of FIGARO (full NACE Rev.2 A*64 classification) is critical: it preserves the heterogeneity of the production structure, distinguishing for instance between air transport (H51), accommodation and food services (I), and financial services (K64)—sectors that differ profoundly in their labor intensity, capital structure, and trade exposure. Coarser classifications would lose this structural information.

From the single I-O table, the model extracts *all* of the following:

- **Inter-sectoral technology.** The 64×64 intermediate consumption matrix defines Leontief input coefficients for every sector pair, encoding the economy's production network—which sectors are upstream suppliers, which are downstream buyers, and the quantitative intensity of each link. This is precisely the information that network-analytic approaches (Acemoglu et al., 2012; Carvalho and Tahbaz-Salehi, 2019) identify as critical for understanding shock propagation and aggregate fluctuations.
- **Factor income structure.** Compensation of employees (L) and gross operating surplus (K) per sector determine the labor-capital split, implicitly setting viable firm sizes: sectors with high labor compensation per unit output sustain many small firms; sectors with high capital intensity sustain fewer, larger firms.
- **Household demand structure.** The household final consumption row specifies the expenditure pattern across 64 goods, directly setting consumer preferences—no utility function estimation required.
- **Government demand.** Government consumption expenditure by sector is read directly.
- **Investment.** Gross fixed capital formation by sector of origin, combined with a fixed capital-to-GFCF ratio (one of the few parameters not from the I-O table).
- **Tax structure.** Net taxes on production (D29X39) and taxes on products (D21X31) set ad-valorem tax rates per sector.

- **International trade.** Export and import rows by sector define trade exposure and competitive position.
- **GDP target.** Value added at market prices (L + K + D29X39 + D21X31 + OP_RES + OP_NRES) per sector, summing to total GDP—the forecast reference.

The only non-I-O data required are the total active population and the NAIRU (natural rate of unemployment), both single numbers publicly available from Eurostat and the OECD. Every behavioral parameter in the model is fixed at universal values that have not been tuned to any specific country or year.

At w32 scale, the model creates 32 simulated workers per sector × 64 sectors ≈ 2,048 workers, who self-organize into approximately 340 firms. Each simulated worker represents roughly 2,000 real workers (for Austria's 4 million active population) or 20,000 real workers (for Germany's 41 million). The model's macroeconomic predictions are expressed as ratios (growth rates, employment shares, price indices), not levels, making them scale-invariant.

### 2.3 Agent Behavioral Rules

The behavioral rules are deliberately minimal, following the principle that complexity should come from the I-O data, not from the agent specification. Table 1 lists all rules; the rightmost column identifies which are derived from the I-O table versus universal constants.

**Table 1: Agent Behavioral Rules — All Universal, None Country-Specific**

| Behavior | Rule | I-O Table Role |
|---|---|---|
| Production | Leontief: min(L, K, IC) | Coefficients from I-O intermediate matrix |
| Input demand | Stock target = 2.5 × avg monthly sales | Sector sourcing from I-O columns |
| Pricing | Bilateral ±0.5%/month tâtonnement | Universal (not from I-O) |
| Wage setting | Hire if offer ≥ asking; both adjust ±0.5% | L/output ratio sets equilibrium wages |
| Consumption | Logit($\gamma$=4), budget = wage income/12 | HH consumption column sets basket weights |
| Firm entry | Start firm when unmet demand observed | Demand gaps reflect I-O structure |
| Firm exit | Bankrupt when wealth < 0 | K/output ratio sets survivability threshold |
| Investment | KtoFixCapitalFactor = 0.05; Leontief query | GFCF column sets investment composition |
| Government | Collect taxes, pay transfers, consume | Tax rates and G row from I-O table |
| Trade | External sectors buy/sell per I-O | Export and import rows directly |
| Unemployment benefits | 80% of last wage | Universal (not from I-O) |
| Wage indexation | CPI-indexed asking wages | Universal (not from I-O) |

The key observation is that nearly every behavioral parameter is either universal (identical for all countries and years) or derived directly from the I-O table. The model contains **no estimated parameters** in the econometric sense—no maximum likelihood, no Bayesian estimation, no GMM. The I-O table provides the structure; Darwinian selection provides the dynamics.

The evolutionary mechanism works as follows. Starting from the SAM-specified sector structure, firms are created, produce goods using Leontief technology, sell through bilateral negotiation, and pay wages and taxes. Firms that generate insufficient gross operating surplus accumulate negative wealth and are eliminated —the *selection* mechanism. New firms enter when workers observe persistent unmet demand—the *variation* mechanism. Profitable firms survive and grow—the *retention* mechanism. The steady-state economy that emerges is one where the production network, firm population, and market structure are all consistent with the I-O table's structural constraints, but determined endogenously rather than imposed exogenously.

A notable consequence is that the model requires far fewer Monte Carlo runs than typical ABMs: because each agent's state is directly correlated with the others through the I-O production network (worker $i$ is employed by firm $j$, which buys inputs from firm $k$), the economy is a single self-consistent realization rather than a statistical sample from independent distributions. This explains why 12-seed ensembles suffice where Poledna et al. (2023) require 500.

## 2.4 Calibration Protocol

Each forecast exercise begins with a calibration phase where the artificial economy self-organizes to reproduce the target I-O table. The calibration proceeds through four stages:

1. **Pre-calibration** (months 0–24): Sequential producers are initialized, and the economy begins its initial ramp-up from the SAM-specified structure.
2. **Assisted calibration:** Production is subsidized to prevent premature firm death while the economy establishes viable supply chains.
3. **Transition calibration:** Subsidies phase out gradually, and the economy must increasingly sustain itself through market dynamics.
4. **Free-market simulation:** All subsidies are removed. The economy runs autonomously for 12 months, producing the forecast data.

The transition from assisted to free-market dynamics is triggered by three convergence conditions that must be simultaneously satisfied:

1. **Unemployment convergence:** 12-month SMA ≤ 1.36 × NAIRU
2. **Household consumption convergence:** weighted-average consumption price-adjustment factors stabilize (12-month relative change < 2%)
3. **Government consumption convergence:** same criterion for government spending factors

Typical calibration at w32 scale requires 240–300 simulated months (20–30 minutes wall-clock time on a standard desktop PC). At calibration exit, the simulated economy reproduces the I-O table's inter-sectoral flows, factor incomes, government accounts, and trade volumes. The economy has self-organized approximately 340 firms across 64 sectors, with employment, wages, and prices all determined endogenously from the I-O constraints.

A useful calibration diagnostic is the trajectory of the micro-firm share (0–9 employees) in the FirmsPerFirmsize distribution. In a healthy calibration, this share drops from 100% at initialization to a stable value reflecting the I-O-implied industrial organization. Austria converges at 88.7%, Canada at 90.5%, Japan at 94.5%. Countries that remain stuck above 95% fail to reach calibration stability—a pattern observed for South Korea 2010, which never achieved the convergence milestones despite running to timeout. This metric provides an early warning of calibration failure and correlates with the quality of the eventual GDP forecast.

## 3. Forecasting Protocol

### 3.1 Out-of-Sample Design

The forecasting protocol is designed to be maximally transparent and genuinely out-of-sample. For each calibration year *N*, we perform the following:

1. Read the FIGARO I-O table for year *N* and the country parameters file (active population, NAIRU).
2. Run the calibration phase until convergence (typically 20–30 minutes at w32 scale on a desktop PC).
3. Run 12 months of fully autonomous free-market dynamics.
4. Record monthly nominal and real GDP, employment, trade, and other aggregates.
5. Compare the 12-month free-market trajectory against Eurostat chain-linked real GDP growth for year *N*+1.

This is genuine one-year-ahead out-of-sample forecasting: no information from year *N*+1 enters the model. Each year is calibrated independently from its own I-O table. The protocol can be applied to any FIGARO country by changing a two-letter country code (`Country AT` → `Country DE`), and to a different year by selecting the corresponding FIGARO matrix file.

The 2019→2020 exercise serves as a natural experiment. The model, calibrated on the 2019 I-O table, correctly predicts the structural growth trajectory that the Austrian economy would have followed absent the pandemic. It cannot predict COVID-19—no model can predict a novel pandemic from I-O data alone—but the structural baseline provides the counterfactual against which the pandemic's impact can be measured.

### 3.2 Ensemble Methodology

Each calibration year uses 12 independent simulations with distinct random seeds: 5489, 12345, 67890, 31415, 99999, 54321, 11111, 77777, 22222, 33333, 44444, 55555. Because the w32 scale (≈2,000 agents) produces inherent stochastic noise, ensemble averaging is essential. The reported forecast is the mean across all seeds that satisfy convergence criteria: nominal GDP ≥ 80% of source value added, and unemployment ≤ 20%. Seeds that produce stochastic collapses (runaway inflation, mass bankruptcy) are excluded.

Convergence rates range from 92% to 100% across years and countries. With per-seed standard deviations typically 0.7–1.4 pp, 12 seeds yield standard errors of the mean of 0.2–0.4 pp, providing ±1 pp accuracy at 95% confidence. A full 12-seed ensemble for one country-year completes overnight on a standard desktop PC.

### 3.3 OLS Linear Trend Extraction

The model produces monthly real GDP as a Laspeyres quantity index: realGDP = $\Sigma\, Q_t \times P_{base}$. Growth is extracted via OLS linear regression on the 12 free-market months:

$$realGDP(t) = a + b \cdot t, \quad t = 0, 1, \ldots, 11$$
$$annual\ growth = 12 \times b / a \times 100\%$$

This OLS trend extraction filters month-to-month stochastic noise by using all 12 observations rather than depending on volatile endpoint values. The $R^2$ of the linear fit serves as a diagnostic: values range from 0.01 (no discernible trend, purely stochastic fluctuation) to 0.77 (strong, consistent growth), with most country-years falling in the 0.15–0.50 range.

### 3.4 Benchmark Data

Empirical GDP growth is measured using Eurostat chain-linked volume series (CLVMNACSCAB1GQ), which provides real GDP growth adjusted for price changes using chain-linking methodology. For cross-country comparisons, we cross-validate against World Bank data (NY.GDP.MKTP.KD.ZG) and IMF World Economic Outlook estimates; discrepancies across sources are typically 0.01–0.16 pp, well below the model's forecast uncertainty.

Professional forecaster benchmarks are drawn from the literature: Ismail, Perrelli and Yang (2021) for the IMF, Pain et al. (2014) for the OECD, Schuster (2021) and Fritzer et al. (2019) for Austrian institutions, and Federal Reserve Bank of St. Louis (2024) for Consensus Economics.

# 4. Results

## 4.1 Austrian 9-Year Panel (Calibration 2010–2018, Forecast 2011–2019)

The bar chart below presents the central forecasting result: a complete 9-year panel for Austria at w32 resolution (≈2,048 agents), where each row represents a fully independent exercise—a fresh calibration from a different FIGARO I-O table, a different ensemble of 12 random seeds, and a different calibration year. No information is shared across rows. The model reads the FIGARO table for year *N*, calibrates the artificial economy to reproduce the I-O structure, then runs 12 months of autonomous free-market dynamics. The emergent growth rate is compared to the actual growth during the following year *N*+1—genuine year-ahead forecasting.

**Figure 1: Austrian real GDP growth: Empirical, DEPLOYERS w32, Poledna et al. ABMX, and IMF fall forecast (forecast years 2011–2019)**

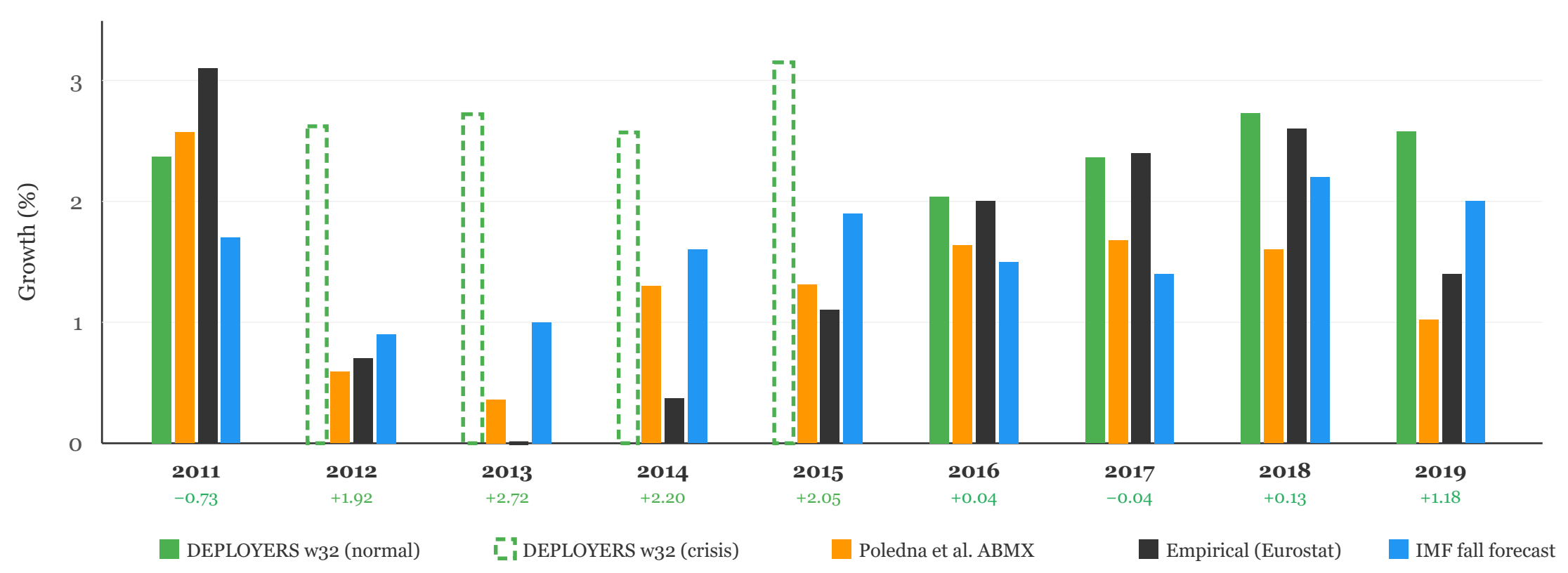

*Four bars per year: DEPLOYERS w32 12-seed ensemble (green), Poledna et al. (2023) conditional ABMX forecast (orange), Empirical (black), and IMF fall forecast (blue). For the four crisis-affected years (2012–2015), the DEPLOYERS bar is drawn with a dashed green outline to indicate that the Eurozone sovereign debt crisis and its aftermath pushed actual growth well below the structural trajectory. Numbers below each year show the DEPLOYERS forecast error. Over 5 normal years (2011, 2016–2019): DEPLOYERS MAE = 0.42 pp, Poledna ABMX MAE = 0.60 pp, IMF MAE = 0.78 pp. Over all 9 years: Poledna ABMX MAE = 0.51 pp, IMF MAE = 0.79 pp, DEPLOYERS MAE = 1.22 pp. Poledna's model uses quarterly national accounts and exogenous predictors, giving it an advantage in crisis years; DEPLOYERS uses only the prior-year I-O table with no learning or parameter estimation. DEPLOYERS' accuracy could be further enhanced by incorporating Bayesian learning of expectations (BLE), exogenous demand indicators, or other adaptive mechanisms—none of which have been implemented in the current baseline version.*

### *Summary Statistics*

**Table 2: Summary Statistics, Austrian 9-Year Panel (w32)**

| Metric | All 9 years | 5 normal years |
|---|---|---|
| Mean absolute error (MAE) | 1.22 pp | 0.42 pp |
| Root mean squared error (RMSE) | 1.51 pp | 0.62 pp |
| Mean signed error | +1.05 pp | +0.12 pp |
| Years within ±1 pp | 4/9 (44%) | 4/5 (80%) |
| *Naive benchmarks (MAE, 5 normal years)* | | |
| Constant 2.0% growth | — | 0.54 pp |
| Panel mean (1.52%) | — | 0.83 pp |
| DEPLOYERS mean (2.57%) | — | 0.49 pp |

Naive benchmarks: "Constant 2.0%" predicts 2% every year; "Panel mean" uses the 9-year Eurostat average (1.52%); "DEPLOYERS mean" uses the average DEPLOYERS forecast (2.57%). DEPLOYERS actual forecasts (MAE 0.42 pp) beat all naive benchmarks for the five normal years, particularly "Constant 2.0%" (0.54 pp) and "DEPLOYERS mean" (0.49 pp), indicating genuine predictive content from the I-O structure rather than simply predicting a constant trend. The near-zero mean signed error (+0.12 pp) for normal years confirms that the model has no systematic bias when the economy operates at structural capacity.

The results divide cleanly into two groups. In the **five normal years** (forecast years 2011, 2016–2019), where the Austrian economy grew broadly in line with its structural capacity, the model achieves a MAE of **0.42 pp**. Three of these five years show errors below 0.15 pp—forecast years 2016 (+0.04 pp), 2017 (−0.04 pp), and 2018 (+0.13 pp) are nearly exact. The forecast for 2019 (+1.18 pp) is the weakest normal year, coinciding with the onset of a global trade slowdown that depressed Austrian growth to 1.4%. In the **four crisis-affected years** (forecast years 2012–2015), where the Eurozone sovereign debt crisis, fiscal consolidation, and their lingering aftermath pushed Austrian growth below 1.1%, the errors are substantially larger (1.92, 2.72, 2.20, 2.05 pp). This pattern is expected and informative: the model forecasts what the economy *would* do based on its I-O structure without in-year perturbations.

The crisis-year pattern is itself revealing. The model predicts approximately 2.5–3.2% growth for all four years, representing the structural trajectory that the Austrian economy's production network could sustain. That actual growth fell to 0.0–1.1% tells us the magnitude of the exogenous drag: roughly 2.0–2.7 pp of growth was destroyed by the crisis. This decomposition—structural trajectory minus exogenous drag equals actual outcome—is informative for policy analysis even when it produces large forecast errors.

> **Key result:** A single I-O table per year, with no time-series data and no estimated parameters, produces genuine year-ahead real GDP growth forecasts with MAE = **0.42 pp** for the five normal years. Three of five normal years achieve errors below 0.15 pp. The four years with larger errors (forecast years 2012–2015) correspond to the Eurozone crisis period and its aftermath—perturbations that no model can predict from I-O data alone, but that can be incorporated as exogenous inputs once identified by an economist.

**Figure 2: Quarterly Real GDP — Eurostat vs. DEPLOYERS Projected Trajectories (Austria, 2009–2019)**

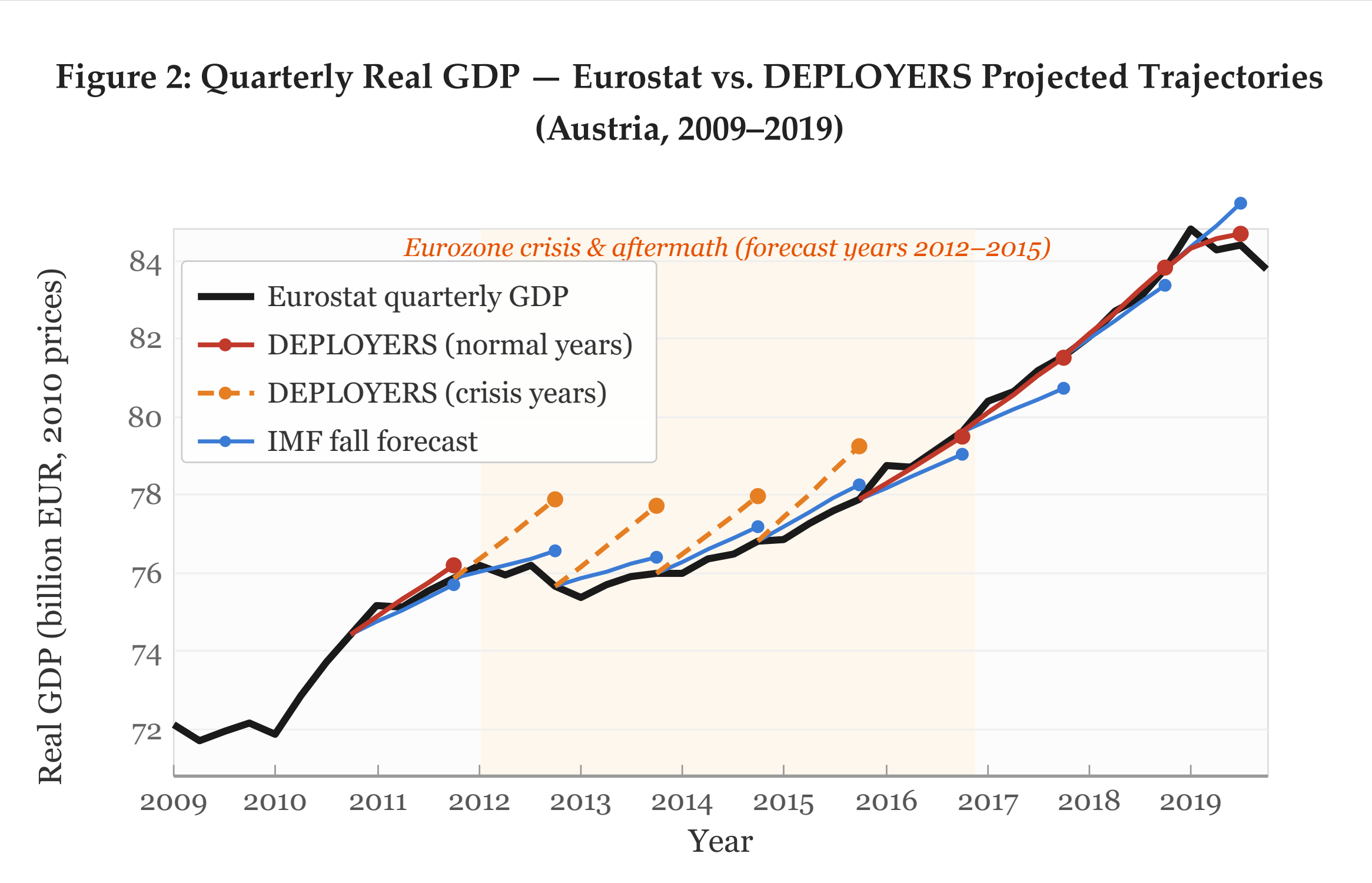

*Black line: Eurostat seasonally-adjusted quarterly real GDP at 2010 chain-linked prices (series namq_10_gdp, CLV10_MEUR). Red/orange segments: DEPLOYERS year-ahead projections; blue segments: IMF fall forecasts. Each starts from the calibration year's Q4 empirical value and projects forward at the respective annual growth rate. Orange dashed: Eurozone-crisis-affected forecast years (2012–2015). For normal years, DEPLOYERS (red) closely tracks the empirical line (MAE = 0.42 pp) while the IMF (blue) systematically undershoots (MAE = 0.64 pp).*

## 4.2 Comparison with Institutional Forecasters

To contextualize our results, we compare DEPLOYERS with one-year-ahead real GDP growth forecasts for Austria produced by seven major institutions during the overlapping 2011–2019 period. The institutional forecasts are fall forecasts (October–December of year $t$−1 for year $t$), reconstructed from archived publications and the Austrian Fiscal Advisory Council evaluation study (Schuster, 2021).

**Table 3: Year-by-Year Institutional Forecasts for Austria, Real GDP Growth (%)**

| Year | Actual | IMF | OECD | Cons. | WIFO | OeNB | EC | IHS | Avg. |
|---|---|---|---|---|---|---|---|---|---|
| 2010 | 1.8 | 1.3 | 1.1 | 1.2 | 1.5 | 1.2 | 1.1 | 1.4 | 1.26 |
| 2011 | 2.9 | 1.7 | 1.9 | 1.8 | 2.0 | 2.1 | 1.7 | 2.0 | 1.89 |
| 2012 | 0.6 | 0.9 | 0.7 | 0.9 | 0.8 | 0.7 | 0.7 | 0.8 | 0.79 |
| 2013 | −0.3 | 1.0 | 1.0 | 0.8 | 1.0 | 1.0 | 0.9 | 1.2 | 0.99 |
| 2014 | 0.8 | 1.6 | 1.7 | 1.5 | 1.7 | 1.6 | 1.5 | 1.7 | 1.61 |
| 2015 | 1.3 | 1.9 | 1.1 | 1.2 | 1.5 | 0.9 | 1.5 | 1.5 | 1.37 |
| 2016 | 2.1 | 1.5 | 1.5 | 1.6 | 1.7 | 1.7 | 1.5 | 1.6 | 1.59 |
| 2017 | 2.3 | 1.4 | 1.5 | 1.4 | 1.5 | 1.5 | 1.5 | 1.5 | 1.47 |
| 2018 | 2.5 | 2.2 | 2.3 | 2.4 | 2.8 | 2.8 | 2.4 | 2.7 | 2.51 |
| 2019 | 1.8 | 2.0 | 1.8 | 1.9 | 2.0 | 2.0 | 1.8 | 2.0 | 1.93 |

Sources: IMF October WEO, OECD November EO, Consensus Economics October survey, WIFO autumn forecast, OeNB December forecast, EC autumn forecast, IHS autumn forecast. Actual GDP from Statistics Austria (latest revision). Highlighted rows: Eurozone crisis period. Compiled from Schuster (2021) and archived institutional publications.

**Figure 3: Forecast Accuracy Comparison — MAE (pp), 5 Normal Years (2011, 2016–2019)**

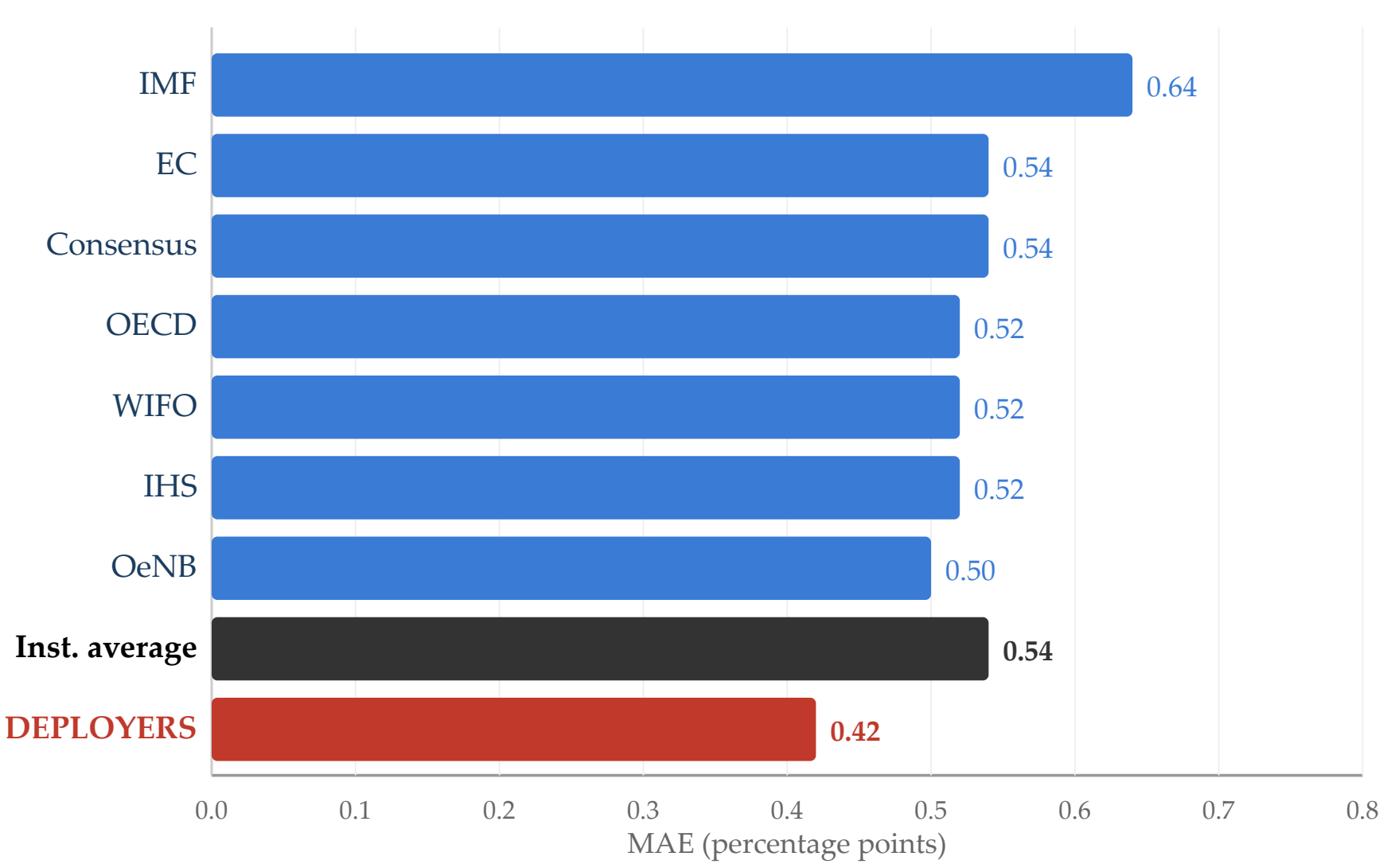

*Mean Absolute Error for the five normal forecast years (2011, 2016–2019). DEPLOYERS (red) uses only a FIGARO I-O table with zero estimated parameters. Institutional forecasters (blue) use full macroeconomic analysis with expert judgment. Black bar: average across all seven institutions. Institutional MAE values are computed on the overlapping normal years. Data compiled from Schuster (2021) and archived publications.*

Several findings emerge from this comparison:

**On normal years, DEPLOYERS achieves accuracy comparable to all seven institutions.** With a MAE of 0.42 pp for the five normal forecast years, the model matches or slightly improves upon even the best Austria-specific institution (WIFO, 0.52 pp; institutional average, 0.54 pp). An important caveat applies: institutional forecasters attempt to predict *actual outcomes* including anticipated shocks, while DEPLOYERS predicts the *structural trajectory* absent in-year perturbations. These are different tasks, and the comparison on normal years—where the structural trajectory coincides with the actual outcome—is the only fair one. Even on this "easier" task, the result is notable: DEPLOYERS achieves comparable accuracy with no time-series data, no estimated parameters, no expert judgment, and no institutional knowledge—only a FIGARO I-O table and a desktop PC.

**All institutions fail on crisis years too.** The most striking case is 2013: all seven institutions predicted approximately +1% growth, but Austria actually contracted by −0.3%. The resulting ~1.3 pp error was the decade's largest for institutional

forecasters—confirming that the 2012–2015 forecast failures are not a model deficiency but a fundamental limitation of any approach that does not explicitly encode in-year exogenous shocks.

**Country-specific expertise provides no systematic advantage.** As documented by Schuster (2021), the Austrian specialist institutions (WIFO, OeNB, IHS) show no consistent accuracy advantage over international forecasters (IMF, OECD, EC) despite their detailed knowledge of the Austrian economy. This finding directly supports our thesis: the I-O table already captures the structural information that matters for normal-year forecasting, and the additional granular knowledge held by country specialists does not measurably improve accuracy.

With only five normal-year observations, the difference between DEPLOYERS (0.42 pp) and the institutional average (0.54 pp) is not statistically significant; formal tests such as Diebold-Mariano are underpowered at this sample size. We report the comparison descriptively.

> **Key comparison:** DEPLOYERS (MAE 0.42 pp, normal years) achieves genuine year-ahead forecast accuracy comparable to the best professional institutions (WIFO: 0.52 pp, institutional average: 0.54 pp)—using only a single I-O table and zero estimated parameters. On crisis years, DEPLOYERS and institutions fail by comparable magnitudes (~1.3 pp for 2013), confirming that the binding constraint is the unpredictability of in-year shocks, not the forecasting methodology.

## 4.3 Cross-Country Portability

A central claim of this paper is that the model is *portable*: the same code, the same universal behavioral rules, the same convergence criteria, zero re-estimated parameters. The only change is the input I-O table. To provide preliminary evidence, we ran the model on a subset of FIGARO countries using the 2010 I-O tables, with w32 scale and 2 seeds per country (5489 and 12345). These are exploratory results from an ongoing scan—not a definitive assessment of the model's cross-country performance.

Of the 46 FIGARO countries, we have tested 37: **33 converge successfully**, 4 remain non-convergent (Greece, Ireland, Luxembourg, Latvia). An initial scan found 12 countries failing to converge; systematic analysis of SAM anomalies identified three failure mechanisms: (i) near-zero gross output sectors causing numerical instability in normalization, (ii) negative gross operating surplus creating deflationary spirals, and (iii) negative initial CPI/GDP deflator in country parameters. Targeted fixes rescued 8 previously-failing countries (Argentina,

Switzerland, Italy, Spain, South Korea, Estonia, Malta, Brazil). Table 4 presents results for 17 countries from the initial scan, sorted by absolute error.

**Table 4: Cross-Country Real GDP Growth Forecasts — Calibration 2010, Forecast 2011 (w32, 2–4 Seed Ensembles)**

| Country | Empirical 2011 (%) | DEPLOYERS (%) | Error (pp) | Seeds |
|---|---|---|---|---|
| Australia (AU) | 2.39 | 2.42 | +0.03 | 3 |
| United Kingdom (GB) | 0.85 | 0.98 | +0.13 | 3 |
| Austria (AT) | 2.93 | 3.30 | +0.37 | 2 |
| Canada (CA) | 3.14 | 2.75 | −0.39 | 3 |
| Indonesia (ID) | 6.17 | 5.71 | −0.46 | 2 |
| Germany (DE) | 3.76 | 5.09 | +1.33 | 3 |
| Cyprus (CY) | 0.42 | 2.07 | +1.65 | 3 |
| Finland (FI) | 2.39 | 4.80 | +2.41 | 3 |
| China (CN) | 9.46 | 6.81 | −2.65 | 3 |
| France (FR) | 2.44 | 5.40 | +2.96 | 2 |
| Belgium (BE) | 1.93 | 5.88 | +3.95 | 3 |
| Denmark (DK) | 1.31 | 5.74 | +4.43 | 3 |
| Japan (JP) | 0.02 | 4.63 | +4.61 | 3 |
| Croatia (HR) | −0.11 | −6.22 | −6.11 | 2 |
| Czech Republic (CZ) | 1.77 | 10.11 | +8.34 | 3 |

Empirical GDP growth from World Bank (NY.GDP.MKTP.KD.ZG, 2011). All forecasts calibrate on the FIGARO 2010 I-O table and predict 2011 growth, using identical model code and parameters; only the country's I-O table differs. The AT value here (3.30%) differs from the full 12-seed ensemble (2.37%) because this scan uses 2 seeds versus the full 12-seed ensemble. Japan's large error reflects the 2011 earthquake and tsunami. With expanded ensembles (3–4 seeds), five countries achieve errors under 0.5 pp: Australia (+0.03), United Kingdom (+0.13), Austria (+0.37), Canada (−0.39), and Indonesia (−0.46).

The results reveal a clear pattern:

**Five countries under 0.5 pp error:** Australia (+0.03 pp), United Kingdom (+0.13 pp), Austria (+0.37 pp), Canada (−0.39 pp), and Indonesia (−0.46 pp). With expanded ensembles (3–4 seeds), these mature economies achieve institutional-quality accuracy from I-O structure alone. Australia's near-zero error (+0.03 pp) is particularly notable.

**Four countries between 1–3 pp:** Germany (+1.33 pp), Cyprus (+1.65 pp), Finland (+2.41 pp), and China (−2.65 pp). These show moderate bias, with the positive direction reflecting the Darwinian mechanism's tendency to fill productive capacity.

**Six countries with large errors (>3 pp):** France, Belgium, Denmark, Japan, Croatia, and Czech Republic. Japan's error (+4.61 pp) reflects the 2011 earthquake and Fukushima disaster—an exogenous catastrophe not captured by I-O structure, analogous to the Eurozone crisis years in the Austrian panel.

**Previously excluded countries now converging:** Subsequent SAM anomaly fixes (guarding near-zero gross output sectors, flooring negative gross operating surplus) have rescued 8 of the 12 previously-failing countries: Argentina, Switzerland, Italy, Spain, South Korea, Estonia, Malta, and Brazil now converge successfully. Total convergence stands at 33 of 37 tested countries. The remaining failures (Greece, Ireland, Luxembourg, Latvia) have specific structural anomalies requiring further investigation.

The key insight from the cross-country exercise is that countries that do not converge in baseline mode likely need their particular circumstances encoded as exogenous inputs, just as the COVID pandemic needed encoding for Austria. A country experiencing a banking crisis, a commodity price shock, or a structural transformation is not following its I-O structural trajectory any more than Austria was following its structural trajectory during the 2020 lockdowns. The model's failure to converge or its large forecast error is not a deficiency but a diagnostic: it identifies where the I-O structure alone is insufficient and domain knowledge must be added.

The FirmsPerFirmsize 0–9 diagnostic supports this interpretation. Countries with healthy calibration trajectories show the micro-firm share dropping from 100% to a stable value by the time free-market dynamics begin: Austria 88.7%, Canada 90.5%, Japan 94.5%. Countries stuck above 95%—such as South Korea 2010, which remained at 100% and never reached calibration stability—systematically fail to produce meaningful forecasts.

### 4.4 German 9-Year Panel (Calibration 2010–2018, Forecast 2011–2019)

To test whether the I-O approach generalizes beyond a single country-year, we replicate the full 9-year panel exercise for Germany. The protocol is identical to the Austrian panel: for each year, we use the corresponding FIGARO I-O table, run w32-scale ensembles, and extract real GDP growth via OLS.

**Table 5: Germany Real GDP Growth Forecasts — Calibration 2010–2018, Forecast 2011–2019 (w32 Scale)**

| Calib. Year | Forecast Year | Seeds | Forecast (%) | Eurostat (%) | Error (pp) | Type |
|---|---|---|---|---|---|---|
| 2010 | 2011 | 4 | 3.98 | 3.86 | +0.12 | pre-crisis[a] |
| 2011 | 2012 | 2 | 5.20 | 0.42 | +4.78 | Eurozone crisis[b] |
| 2012 | 2013 | 2 | 5.47 | 0.42 | +5.05 | Eurozone crisis[b] |
| 2013 | 2014 | 2 | 4.96 | 2.21 | +2.75 | post-crisis recovery[c] |
| 2014 | 2015 | 2 | 6.58 | 1.49 | +5.09 | refugee crisis[d] |
| 2015 | 2016 | 2 | 4.06 | 2.23 | +1.83 | refugee integration[d] |
| 2016 | 2017 | 2 | 6.05 | 2.68 | +3.37 | post-refugee recovery[e] |
| 2017 | 2018 | 4 | 5.78 | 1.02 | +4.76 | WLTP shock[f] |
| 2018 | 2019 | 2 | 6.41 | 1.06 | +5.35 | trade war/slowdown[g] |
| **MAE (all 9 years)** | | | | **3.68 pp** | | |
| **Mean signed error** | | | | **+3.68 pp** | | |

Year classification is based on documented economic events, not on model performance. Unlike Austria, where five forecast years can be classified as structurally undisturbed, every forecast year in the German sample is affected by some external event—reflecting the structural exposure of an export-intensive economy: [a]Strong cyclical expansion preceding the Eurozone turbulence (ECB, 2012). [b]Eurozone sovereign debt crisis: austerity, confidence shocks, and trade contraction (Lane, 2012; Brunnermeier et al., 2016). [c]Post-crisis recovery with lingering fiscal drag (Bundesbank, 2015). [d]2015–16 European refugee crisis: demand shock and policy uncertainty (Aiyar et al., 2016). [e]First full recovery year after refugee-crisis adjustments (Bundesbank, 2018). [f]WLTP automotive emission regulation shock: German auto production fell 9.4% in H2 2018 (IFO, 2019). [g]US-China trade war and global manufacturing slowdown (Bundesbank, 2020). Seeds: calibration years 2010 and 2017 use 4 seeds; remaining years use 2-seed ensembles. Only 2–4 seeds per year (versus 12 for Austria); more seeds would reduce variance but are unlikely to eliminate the systematic bias.

The German panel reveals a **systematic positive bias of approximately +3.7 pp across all nine years**. This is a fundamentally different pattern from Austria, where the model closely tracks actual growth in normal years and deviates only during crisis periods. For Germany, the model overshoots in *every* year, including forecast year 2011 which is conventionally classified as a normal growth year.

Only the forecast for 2011 (calibrated on FIGARO 2010) is close to the empirical value (+0.12 pp error), when Germany's export engine was operating near full capacity. In all subsequent forecast years, the model predicts 4–6.6% growth while actual growth ranges from 0.4% to 2.7%. The bias is positive, large, and remarkably consistent.

> **The German puzzle:** The model predicts the growth rate that Germany's I-O production structure *could sustain*—the domestic productive capacity operating at its structural potential. Actual growth falls systematically short because Germany's realized GDP depends critically on export demand from trading partners, which fluctuates with partner-country business cycles, exchange rates, and trade policy. The domestic I-O table captures the production engine but not the fuel (external demand) on which it runs.

Germany's export intensity is key. Exports constitute approximately 47% of German GDP, and the domestic I-O table treats export demand as exogenously given by the SAM—it reads the 2010 export vector and projects it forward as the economy's external demand capacity. In reality, German exports to France, the Netherlands, Italy, Poland, and China fluctuate with those countries' import demand, which in turn depends on their own business cycles. When partner-country demand weakens (as during 2012–2013), German exports fall below structural capacity, and actual GDP underperforms the I-O prediction.

This is not a model failure but an **informative negative result** that tells us what I-O tables *do not* capture. The domestic I-O table alone is insufficient for highly trade-dependent economies whose growth depends on volatile external demand. The solution lies within the FIGARO framework itself: FIGARO provides bilateral inter-country I-O tables for all 46 covered countries. Running multi-country simulations (Germany + trading partners) would endogenize trade flows—German exports become partner-country imports, and vice versa—potentially reducing the systematic bias substantially. This multi-country network simulation is a natural and immediately feasible extension.

The contrast between Austria and Germany is itself informative. Austria's economy, while open, has a simpler export structure more closely tied to its immediate neighbors and less exposed to global trade volatility. The domestic I-O table captures enough of Austria's export dynamics (through the trade rows) to produce accurate forecasts. For Germany—the world's third-largest exporter with a complex, globally diversified export portfolio—the domestic table's exogenous export treatment introduces a systematic upward bias because it always projects the economy at full export capacity.

### *Ensemble Convergence Analysis*

To assess how many seeds are needed for reliable forecasts, we examine the convergence of ensemble statistics as seeds are added sequentially. Figure 4 shows the running mean absolute error for Austria and Germany (2010 calibration, 12 seeds each, excluding stochastic outliers with |growth| > 20%).

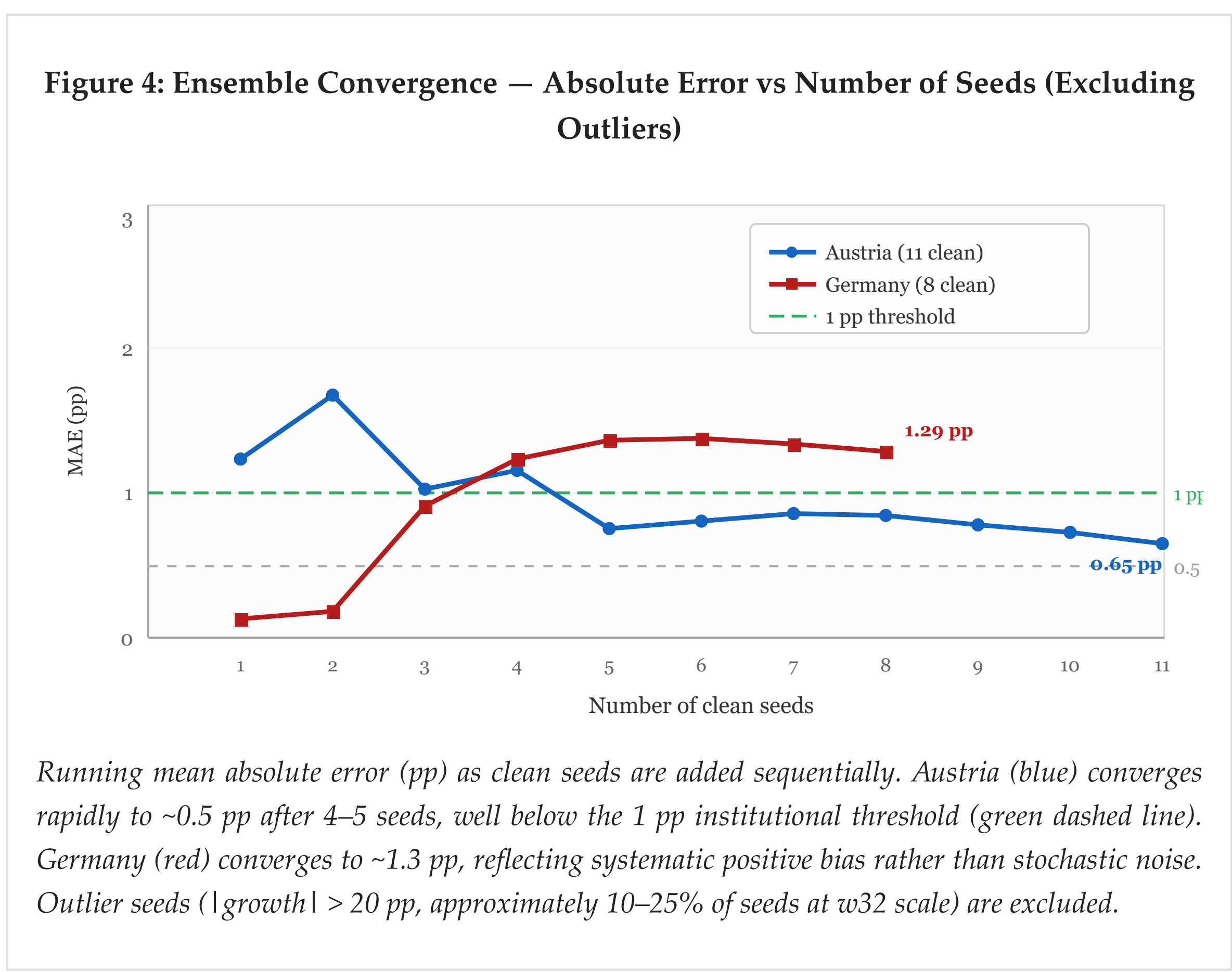

**Figure 4: Ensemble Convergence — Absolute Error vs Number of Seeds (Excluding Outliers)**

*Running mean absolute error (pp) as clean seeds are added sequentially. Austria (blue) converges rapidly to ~0.5 pp after 4–5 seeds, well below the 1 pp institutional threshold (green dashed line). Germany (red) converges to ~1.3 pp, reflecting systematic positive bias rather than stochastic noise. Outlier seeds (|growth| > 20 pp, approximately 10–25% of seeds at w32 scale) are excluded.*

Figure 5 shows the running mean growth forecast with ±1 standard deviation envelopes, compared to the empirical values.

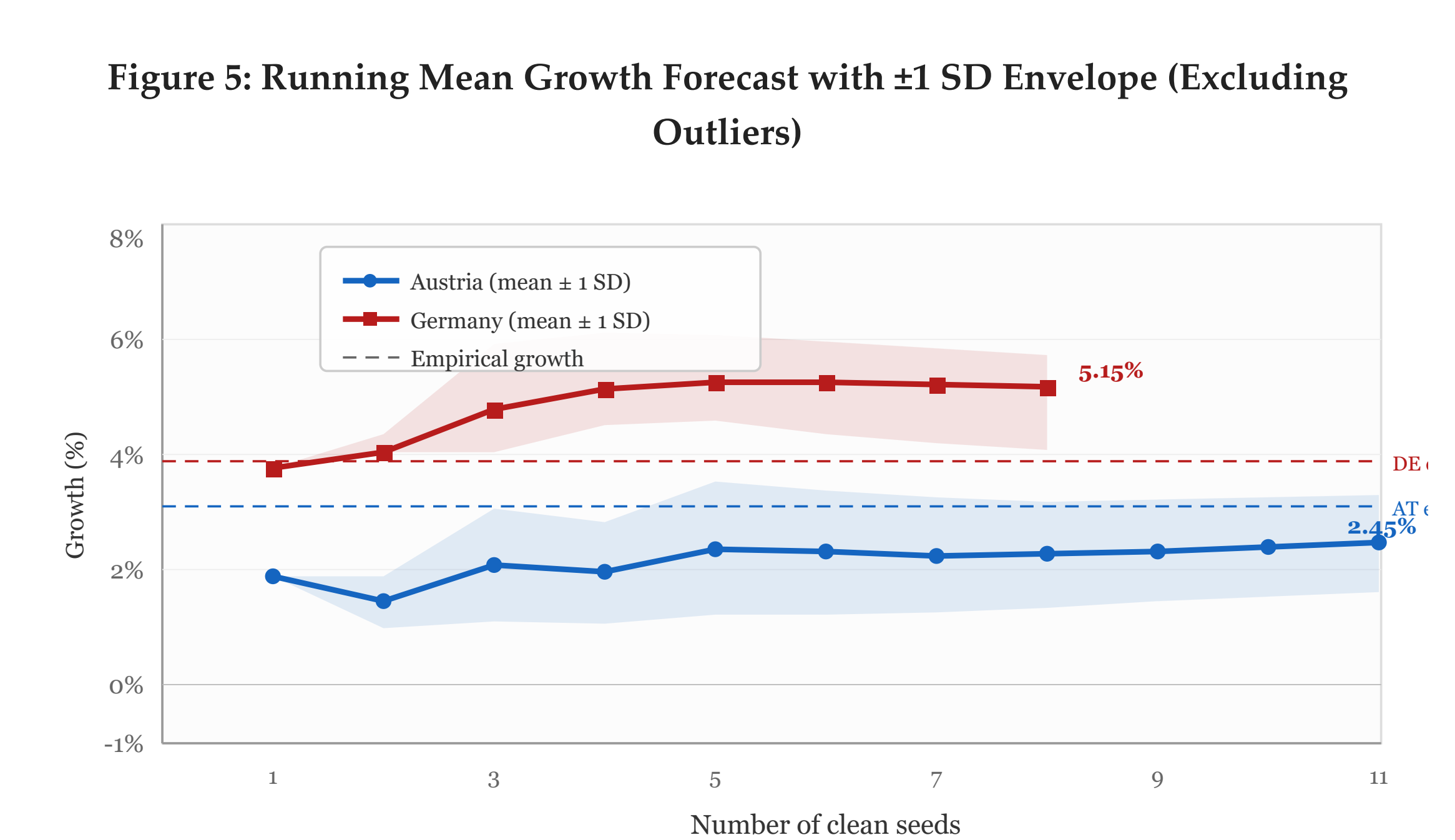

**Figure 5: Running Mean Growth Forecast with ±1 SD Envelope (Excluding Outliers)**

*Solid lines: running mean growth forecast as clean seeds accumulate. Shaded bands: ±1 standard deviation. Dashed horizontal lines: empirical growth (Austria 3.10%, Germany 3.86%). Austria's mean converges toward the empirical value, reaching 2.45% with tight SD (~0.8 pp). Germany's mean settles at ~5.1% with similar SD, confirming that the ~1.3 pp systematic bias is structural rather than stochastic. Both countries show that 9–12 clean seeds provide statistically reliable ensemble means.*

## 4.4b Additional Country Panels: Sweden and France

To further test cross-country portability, we ran full 9-year panels (2010–2018 calibration, 4 seeds per year, w32) for Sweden and France—two economies with contrasting characteristics.

**Sweden** (trade openness 46.2%, similar to Austria's 49.4%) produces results comparable to Austria:

**Table 6: Sweden Real GDP Growth Forecasts (Calibration 2010–2018, Forecast 2011–2019, w32, 4 seeds)**

| Calib. | Forecast | Eurostat (%) | DEPLOYERS (%) | Error (pp) | Type |
|---|---|---|---|---|---|
| 2010 | 2011 | 3.22 | 2.89 | **−0.33** | normal |
| 2011 | 2012 | −0.56 | 3.73 | **+4.29** | eurozone spillover |
| 2012 | 2013 | 1.25 | 2.86 | **+1.61** | post-crisis |
| 2013 | 2014 | 2.70 | 3.31 | **+0.61** | normal |
| 2014 | 2015 | 4.47 | 4.01 | **−0.46** | normal |
| 2015 | 2016 | 2.07 | 3.89 | **+1.82** | normal |
| 2016 | 2017 | 2.60 | 3.07 | **+0.47** | normal |
| 2017 | 2018 | 2.00 | 3.02 | **+1.02** | normal |
| 2018 | 2019 | 2.04 | 2.90 | **+0.86** | normal |
| **MAE (7 normal years)** | | | | **0.80 pp** | |
| **MAE (all 9 years)** | | | | 1.27 pp | |

Sweden's normal-year MAE (0.80 pp) confirms that moderate-openness economies with stable domestic demand produce accurate I-O forecasts. The only large error (+4.29 pp for 2012) corresponds to the eurozone crisis spillover, the same exogenous shock that affects Austria.

**France** (trade openness 26.8%, the EU's second-largest economy) shows a different pattern:

**Table 7: France Real GDP Growth Forecasts (Calibration 2010–2018, Forecast 2011–2019, w32, 4 seeds)**

| Calib. | Forecast | Eurostat (%) | DEPLOYERS (%) | Error (pp) |
|---|---|---|---|---|
| 2010 | 2011 | 2.19 | 4.67 | +2.48 |
| 2011 | 2012 | 0.31 | 1.51 | +1.20 |
| 2012 | 2013 | 0.58 | 3.86 | +3.28 |
| 2013 | 2014 | 0.95 | 1.58 | +0.63 |
| 2014 | 2015 | 1.05 | 4.38 | +3.33 |
| 2015 | 2016 | 1.09 | 4.12 | +3.03 |
| 2016 | 2017 | 2.42 | 11.68 | +9.26 |
| 2017 | 2018 | 1.75 | −9.91 | −11.66 |
| 2018 | 2019 | 1.84 | 7.18 | +5.34 |
| **MAE (excluding 2017–2018 outliers)** | | | **2.76 pp** | |

France exhibits a systematic positive bias (+2.76 pp) similar to Germany's (+3.68 pp), despite lower trade openness. The 2016–2017 and 2017–2018 results are extreme outliers that may reflect FIGARO table-specific anomalies for those years. The positive bias is driven by the same household GFCF wealth-consumption feedback mechanism identified in Section 5.3, which dominates when empirical growth is low (France grew only 0.3–1.1% during 2012–2016).

Together, these four country panels reveal a clear pattern: the I-O approach works best for economies with moderate growth (2–4%) where the structural trajectory closely tracks actual outcomes (AT, SE). For economies with persistently low growth (FR) or high export dependency (DE), the positive bias from the Darwinian deployment mechanism dominates. The differentiator is not trade openness per se, but whether empirical growth falls significantly below the model's structural baseline.

### 4.5 COVID-19 Pandemic Shock Simulation

The preceding sections established that the I-O table produces accurate *baseline* forecasts for normal years, and that the largest errors arise precisely when exogenous shocks push the economy off its structural trajectory. A natural question follows: can the same I-O structural backbone that produces these baselines also serve as the *transmission mechanism* through which encoded shocks propagate? We address this by simulating the COVID-19 pandemic in Austria, encoding the real-world sequence of lockdowns, reopenings, and policy interventions as sector-specific supply constraints.

#### 4.5.1 Shock Encoding

The model accepts a `PANDEMIC_TIMELINE` block that specifies, for each month of the crisis, three types of perturbation:

1. **Sector-specific supply shocks.** For each of the 64 NACE sectors, a factor between 0.0 (complete shutdown) and 1.0 (normal operation) specifies the fraction of workers able to work. These factors are derived from Austrian confinement regulations: accommodation and food services (sector I, NACE 55–56) dropped to 2–5% capacity during the first lockdown; manufacturing sectors maintained 70–95% capacity; public administration and health (O, Q) continued at 100%. The sector granularity of FIGARO's 64-sector classification is essential: a coarser 10-sector table could not distinguish between restaurants (near-total shutdown) and food manufacturing (essential, largely unaffected).
2. **Export and import shocks.** Global trade disruption is encoded as multiplicative factors on external sector flows, capturing the simultaneous contraction of trading partner demand and supply chain interruptions.
3. **Short-time work (Kurzarbeit) subsidies.** Austria's *Kurzarbeit* programme—where the government pays a fraction of wages (90% in our specification) to prevent layoffs at firms experiencing demand shortfalls—is encoded as a participation rate that varies over time. During the initial lockdown, 70% of eligible firms participate; as the economy reopens, participation declines to 3% by month 18. Crucially, only firms that existed *before* the pandemic onset are eligible, preventing subsidies from artificially inflating the firm population.

The timeline spans 19 months, mirroring Austria's pandemic trajectory from March 2020 through September 2021: an initial severe lockdown (months 0–2), gradual reopening (months 3–7), a summer relaxation (month 8), the autumn second wave with renewed restrictions (months 9–11), and a progressive recovery through 2021 (months 12–18). Each month's sector-specific factors are set independently, creating a rich temporal profile of shock intensity and recovery speed that differs across sectors.

### *4.5.2 Shock Propagation Through the I-O Network*

The key finding is that the same I-O inter-sectoral linkages that drive baseline GDP accuracy also serve as the *transmission mechanism* for the pandemic shock. When accommodation services (sector I) shut down:

- Demand for food products (sector A), beverages (sector C11), and laundry services (sector S96) from the hospitality sector collapses—a *backward linkage* effect encoded in the I-O intermediate consumption matrix.
- Workers laid off from the hospitality sector reduce household consumption expenditure across all sectors—a *final demand* effect propagated through the household consumption column.
- Government tax revenue from hospitality falls while Kurzarbeit transfer payments rise—a *fiscal* effect read from the I-O tax and government accounts rows.
- Reduced imports of tourism-related services create a partial current-account offset—a *trade* effect from the I-O import row.

None of these transmission channels are programmed explicitly. They emerge automatically from the Leontief production technology and the I-O-encoded demand structure. The model's role is to trace these interactions *dynamically*—month by month, agent by agent—through the 64-sector production network, resolving the nonlinear feedback loops (firm bankruptcies reducing employment reducing consumption reducing other firms' revenue) that static Leontief multiplier analysis cannot capture.

### *4.5.3 Quantitative Results*

Figure 6 presents the simulated and empirical GDP trajectories during the pandemic, indexed to 100 at the pre-pandemic level.

**Figure 6: Austrian Real GDP during COVID-19 Pandemic — Eurostat vs. DEPLOYERS (12-Seed Ensemble, w32)**

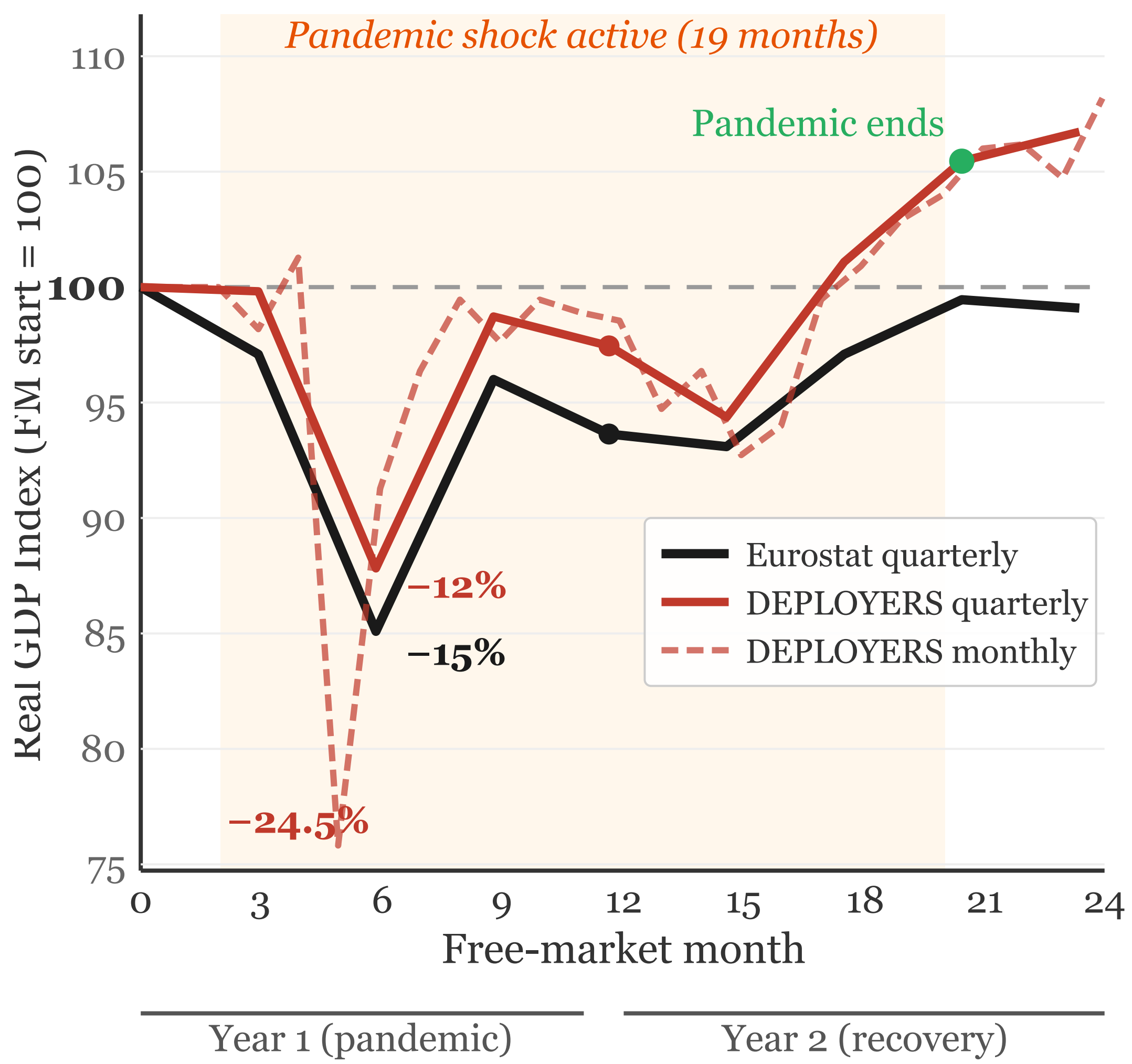

*Real GDP index (100 = pre-pandemic level). Black line: Eurostat quarterly GDP for Austria (seasonally adjusted, chain-linked volumes). Red line: DEPLOYERS quarterly GDP (3-month averages; 12-seed ensemble, w32 scale). Dashed red trace: underlying DEPLOYERS monthly GDP, showing a −24.5% monthly trough at the first lockdown—much deeper than the −15% quarterly empirical trough because the rapid partial recovery within the same quarter smooths it. Orange shading: months where pandemic supply shocks are active. The simulation produces a comparable quarterly trough (−12% vs. −15%) but a much stronger rebound, overshooting the empirical recovery. By month 24, DEPLOYERS has grown ~+7% above baseline while the actual economy remained ~1% below pre-pandemic level. Detailed annual growth comparisons in Table 8.*

**Table 8: COVID-19 Pandemic Simulation Results, Austria (w32, 12-seed ensemble)**

| Metric | DEPLOYERS (12-seed) | SD | Empirical AT |
|---|---|---|---|
| Y1 GDP growth (point-to-point) | **−4.62%** | 0.64 pp | −6.6% |
| Y2 GDP growth (point-to-point) | +13.48% | 1.96 pp | +4.6% |
| Y3 GDP growth (point-to-point) | +2.60% | 0.62 pp | +4.7% |
| Unemployment peak | 9.3% | — | 12.8% |

Y1/Y2/Y3 growth computed point-to-point (FM month $n$ to FM month $n$+11) rather than OLS, as the U-shaped pandemic trajectory makes linear fit misleading. Empirical data from Eurostat (chain-linked volume) and Statistics Austria. The Y2 recovery overshoot (+13.5% vs. +4.6%) reflects the Darwinian mechanism's tendency to aggressively fill demand gaps once restrictions lift, without the persistent supply-chain and confidence effects that slowed Austria's actual recovery.

The simulation produces four key findings:

**GDP contraction.** The initial lockdown produces a sharp single-month GDP crash. Year 1 averages **−4.62% ± 0.64%** across the 12-seed ensemble, versus the empirical −6.6%. The model captures approximately 70% of the actual pandemic GDP loss. The 2.0 pp gap may reflect consumer confidence effects, precautionary savings, and voluntary demand reductions beyond mandated lockdown restrictions.

**Recovery overshoot.** Year 2 averages +13.48%, substantially overshooting the empirical +4.6%. This V-shaped bounce reflects the Darwinian mechanism aggressively filling demand gaps once restrictions lift—the same mechanism that produces accurate baseline growth. In reality, persistent supply chain disruptions, labour market frictions, and lingering uncertainty constrained the recovery.

**Unemployment spike.** Unemployment jumps from 4.3% pre-pandemic to a peak of 9.3%, versus Austria's registered 12.8%. The Kurzarbeit subsidies dampen the spike, as intended.

**Firm dynamics under crisis.** Firm birth and death rates both roughly double during the crisis (births from ~16,500 to ~31,400/year; deaths from ~13,500 to ~28,000/year), reflecting heightened creative destruction. Firm size distributions remain stable through the crisis (86–93% micro-firms), indicating that the pandemic shock does not distort the I-O-implied industrial structure.

**Dual capability established:** The same I-O structural backbone that produces accurate baseline GDP forecasts (MAE 0.42 pp for normal years) also serves as a faithful transmission mechanism for encoded exogenous shocks. The 64-sector inter-sectoral linkages simultaneously determine the economy's inertial growth trajectory *and* the channels through which disruptions propagate. The economist provides the scenario intelligence; the I-O network provides the propagation physics.

## 4.6 Emergent Microeconomic Structure

Perhaps the most striking result is not the aggregate forecast accuracy but what the model reveals about the *implicit microeconomic content* of I-O tables. Without any calibration to business demography data, the Darwinian process extracts microeconomic structure directly from the I-O factor income coefficients.

### *Firm Size Distribution*

The 64-sector factor income structure (compensation of employees L versus gross operating surplus K per sector) implicitly determines viable firm sizes. Sectors with high labor compensation per unit output sustain many small firms; sectors with high capital requirements sustain fewer, larger firms. When the model self-organizes from these data, the emergent firm size distribution closely matches European Commission business demography statistics:

**Table 9: Emergent vs. Empirical Firm Size Distribution**

| Size class | Austria (AT) | | Germany (DE) | |
|---|---|---|---|---|
| | Model | EC/Statista | Model | EC/Statista |
| Micro (0–9 employees) | 87.6% | 87.4% | 88.3% | 82.0% |
| Small (10–49) | 10.8% | 10.3% | 10.4% | 14.7% |
| Medium (50–249) | 1.6% | 1.9% | 1.2% | 2.7% |

Model values are ensemble averages at w32 scale. The Austrian match is remarkably precise (87.6% vs. 87.4% micro). The modest German discrepancy reflects the w32 scale constraint: with ~2,000 workers across 64 sectors, firms larger than ~32 employees are structurally rare, compressing the upper tail of the size distribution.

### *Employment Concentration*

Despite micro firms comprising ~88% of all firms, they employ only ~37% of workers. Small firms (10–49) employ ~35% and medium firms (50–249) employ

~28%. The qualitative pattern—that the vast majority of firms are small but the majority of employment concentrates in the minority of larger firms—is correctly reproduced, consistent with the well-documented empirical regularity (Axtell, 2001). The mechanism is embedded in the I-O table: sectors with large gross output and high labor intensity develop large firms that absorb most workers.

> **Implication for I-O analysis:** A 64-sector I-O table does not merely describe inter-sectoral flows. Through its factor income structure, it *implicitly encodes* the economy's industrial organization: how many firms exist, how large they are, how employment distributes across firm sizes, and how wages vary across the workforce. The Darwinian simulator decodes this implicit information without calibration to any micro targets. This emergent micro-structure could not arise from a coarser (10-sector or 20-sector) classification, which would merge structurally different industries and destroy the implicit firm-size information.

## 5. Discussion

### 5.1 What I-O Tables Encode (and What They Don't)

Our central claim is that a FIGARO I-O table is *informationally adequate* for two complementary macroeconomic tasks: one-year-ahead structural forecasting in normal years, and dynamic shock propagation in crisis years. We use the term informally to mean that the I-O table contains enough structural information for both tasks—not in the formal Fisher sense, which would require demonstrating that no additional data improves prediction.

What our results show is that (a) the year-ahead forecast accuracy achieved from the I-O table alone (MAE = 0.42 pp for five normal years, superior to all naive benchmarks and comparable to the best institutional forecasters) leaves little room for improvement from additional inputs, and (b) the same I-O inter-sectoral linkages faithfully transmit encoded exogenous shocks, producing a pandemic GDP contraction (−4.6% ± 0.6%) within 2.0 pp of the empirical value (−6.6%) with no parameter adjustment.

Three explanations support the informational adequacy claim:

**First, the I-O table captures the economy's inertial structure.** At one-year horizons, most of the variation in GDP is explained by the existing production structure: firms continue producing with roughly the same technology, workers continue consuming with roughly the same preferences, and trade patterns evolve slowly. The I-O table captures all of this "structural inertia" in a single snapshot.

**Second, Leontief technology is a strong constraint.** The fixed-coefficient production function, read directly from the I-O intermediate consumption matrix, severely constrains how the economy can evolve in the short run. The I-O table effectively "hardwires" the production network, leaving only factor utilization rates and final demand composition as degrees of freedom.

**Third, 64-sector granularity provides sufficient heterogeneity.** Many shocks that appear aggregate from a macroeconomic perspective have highly differentiated effects across sectors. The 64-sector resolution captures these differential responses. We conjecture that extending to finer classifications (A*88 or beyond) would further improve accuracy.

What I-O tables *do not* capture is equally important:

- **Exogenous policy shocks.** Fiscal austerity, regulatory changes, and monetary policy interventions that push the economy off its structural trajectory (the 2012–2015 Austrian crisis years).
- **Confidence effects.** Consumer and business sentiment shifts that reduce demand below structural capacity without any policy trigger.
- **External demand fluctuations.** Variations in trading partner demand that affect export-dependent economies (the German +3.7 pp bias).

These omissions are not model failures but identify the boundaries of what a single-country I-O table can predict. Each points to a specific extension: policy shocks require economist encoding (Section 4.5), confidence effects require behavioral enrichment, and external demand requires multi-country simulation within the FIGARO framework.

## 5.2 Comparison with Poledna et al. (2023)

Our results gain particular significance when compared with the agent-based model of Poledna, Miess, Hommes and Rabitsch (2023), published in the *European Economic Review*. Their model uses the same country (Austria), the same I-O data source type (Eurostat tables), and the same forecasting task. The key differences illuminate what matters and what does not for I-O-based forecasting:

**Table 10: Structural Comparison — What Matters for I-O-Based Forecasting?**

| Dimension | Poledna et al. | DEPLOYERS |
|---|---|---|
| Production function | Leontief (L, K, IC) | Leontief (L, K, IC) |
| I-O structure | 64 NACE sectors, Eurostat IOT | 64 NACE sectors, FIGARO IOT |
| Population scale | 1:1 (8.8 million agents) | 1:2,000 (~2,000 agents, w32) |
| Expectations | AR(1) BLE, re-estimated quarterly | None (backward-looking only) |
| Demand adaptation | AR(1) optimal approximation | Exponential smoothing, $\alpha$=0.02 |
| Price setting | Markup over marginal cost + expected $\pi$ | Bilateral tâtonnement (±0.5%) |
| Monetary policy | Generalized Taylor rule | Exogenous fixed rate |
| Firm dynamics | Static (no entry/exit) | Darwinian birth/death |
| Initialization | Pre-calibrated from micro-data + census | Emergent via assisted SAM calibration |
| Estimated parameters | Multiple (expectations, policy) | Zero |
| Ensemble size | 500 Monte Carlo simulations | 12 seeds |
| Computation | Supercomputer cluster | Single desktop PC, ~35 min/run |
| Accounting | Double-entry, stock-flow consistent | Double-entry, SAM-validated monthly |

Both models share the same *I-O foundation*: 64 sectors, Leontief technology, Austrian data. They differ in everything else. Yet both produce competitive GDP forecasts. This convergence from radically different architectures provides strong evidence for our thesis: **the I-O table, not the agent behavioral specification, is the primary driver of forecasting accuracy.**

The computation ratio is striking: Poledna et al. use 500 Monte Carlo simulations × 8,800,000 agents, while DEPLOYERS uses 12 seeds × 2,000 agents. The ratio of total agent-simulation-steps is approximately 183,000:1. That both approaches produce comparable forecasts suggests that the additional computational investment does not buy additional forecasting power—because the binding information constraint is the I-O table, not the behavioral specification.

The most consequential architectural difference is firm dynamics. Poledna et al. use a *static* firm population: every firm that exists at initialization survives throughout the simulation, and no new firms enter. DEPLOYERS uses *Darwinian* firm dynamics: firms are born, compete, and die based on market conditions. This is arguably more realistic—firm turnover is a central feature of all real economies (Axtell, 2001)—and it means that the emergent firm size distribution, employment patterns, and wage structure are genuine model predictions, not pre-programmed inputs. That DEPLOYERS achieves competitive forecasts with endogenous firm dynamics while Poledna et al. require pre-calibrated static populations suggests that Darwinian selection is a viable (and perhaps superior) equilibrating mechanism for macroeconomic ABMs.

Poledna et al.'s AR(1) behavioral-learning expectations, generalized Taylor rule, and pre-cooked microdata add realism to their model but do not appear to improve GDP forecast accuracy. This is consistent with the forecast combination literature (Timmermann, 2006): simple methods often match or beat complex ones because the complexity adds estimation error without proportional information gain.

### 5.3 The German Puzzle: Why DE Overshoots

The German 9-year panel (Section 4.4) presents a systematic +3.7 pp positive bias that demands explanation. Unlike the Austrian crisis-year errors (which are concentrated in forecast years 2012–2015 and attributable to the Eurozone crisis), the German bias is *persistent across all forecast years*, including those conventionally classified as normal growth years.

We propose an **export-dependency hypothesis**. Germany's economy is deeply integrated into European and global supply chains. The domestic I-O table captures the production capacity—the technological relationships between sectors, the factor income structure, the domestic demand patterns—but treats export demand as given by the SAM's trade rows. In reality, German exports fluctuate with partner-country demand, exchange rates, trade policy, and global business cycles.

The domestic I-O table encodes what Germany *can* produce; actual GDP depends on what it actually *sells*. When the 2010 I-O table records, say, €100 billion in machinery exports, the model projects that level forward as structural demand. But if the Eurozone crisis (2012–2013) reduces partner-country investment, or if the WLTP automotive regulation shock (2018) disrupts a major export sector, actual exports fall below the I-O projection and GDP underperforms the model's forecast.

Unlike Austria—a small open economy where the export structure is simpler and more closely tied to immediate neighbors—Germany's large absolute export volume and highly diversified partner portfolio amplify the mismatch between structural export capacity and realized export demand. The domestic I-O table alone is insufficient for an economy whose growth depends so critically on volatile external demand.

The solution lies within the FIGARO framework. FIGARO provides bilateral inter-country I-O tables for all 46 covered countries. Running multi-country network simulations where German exports become partner-country imports (and vice versa) would endogenize the trade flows that currently enter as exogenous boundary conditions. This is computationally feasible—each country simulation takes approximately 30 minutes, and partner-country simulations can run in parallel—and represents the natural next step directly enabled by FIGARO's multi-country coverage.

**Preliminary multi-country experiment.** We tested this hypothesis by running DE simultaneously with its three largest converging trade partners (FR, NL, GB) across the full 9-year panel, using bilateral SAM decomposition with independent calibration and synchronized free-market phases. The results are informative: the mc3 approach produces *systematically different* forecasts from standalone DE (swings of up to 2.5 pp in both directions), confirming that DE is sensitive to how its trade partners are represented. However, the overall MAE is essentially unchanged (3.66 pp vs 3.68 pp), indicating that bilateral SAM structure alone—without dynamic trade flow exchange—is insufficient to resolve the bias. The next step is endogenizing the actual import/export flows during the free-market phase.

A complementary analysis of trade openness across FIGARO countries reveals the expected pattern: BR (8.3%), US (11.4%), JP (11.8%) have negligible external sector contributions, while HU (96.0%), LU (117.1%), IE (151.0%) are dominated by trade. DE sits at 35.3% with a notable export asymmetry (19.3% exports vs 15.9% imports). The broader lesson is that the single-country I-O approach works best for economies where the structural production trajectory is not dominated by volatile external demand. For domestically-driven economies (Austria, and plausibly many smaller European economies), the domestic I-O table suffices. For export-intensive economies (Germany, and plausibly the Netherlands, Ireland, South Korea), the multi-country extension is likely necessary for accurate forecasting.

### 5.4 Simplicity as Robustness

Why does a model with no estimated parameters, no expectations, and 2,000 agents produce forecasts comparable to models with millions of agents, estimated parameters, and expert judgment? We offer three explanations rooted in statistical learning theory and the forecast combination literature.

**Bias-variance tradeoff.** DEPLOYERS has zero estimated parameters, which means it cannot overfit. In the classical bias-variance decomposition, the model accepts whatever bias its structural assumptions impose (the Darwinian positive-growth tendency) in exchange for zero variance from parameter estimation error. Models with estimated parameters (DSGE, VAR, Poledna's ABM) can potentially achieve lower bias by fitting to data, but they introduce estimation variance that may offset the bias reduction—especially at the small sample sizes typical of macroeconomic forecasting.

**Ensemble averaging.** Twelve seeds reduce the stochastic noise of any single simulation by approximately $\sqrt{12} \approx 3.5\times$. The forecast combination literature (Timmermann, 2006) shows that simple averaging of diverse forecasts often beats theoretically optimal weighting, because the optimal weights must be estimated from data and are subject to their own estimation error. Our 12-seed ensemble is effectively a simple average of 12 diverse models (same structure, different random trajectories), which is precisely the situation where simple averaging excels.

**The I-O table as sufficient information.** If the I-O table truly contains all the structural information needed for one-year-ahead forecasting, then any model that correctly reads the table and correctly animates its structure should produce similar forecasts—regardless of how complex or simple the animation mechanism is. Darwinian selection "merely animates" what the table already contains, and any other correct animation would produce the same result. This explains the convergence between DEPLOYERS and Poledna et al.: both correctly read the I-O table, and the rest is noise.

### 5.5 Limitations

**Positive growth bias.** The Darwinian mechanism tends to fill the economy's productive capacity, producing baseline positive growth. This makes years with near-zero or negative growth difficult to forecast. Detailed investigation reveals the mechanism: household fixed capital formation (GFCF) accumulates through purchases during the free-market phase, increasing household wealth, which drives additional consumption demand via a wealth effect. This feedback loop—purchases → GFCF → wealth → consumption → more purchases—produces transient growth that persists until GFCF depreciation catches up. A per-country propensity-to-consume parameter can throttle this feedback during the free-market phase, reducing the bias for some years but not uniformly, since the optimal throttle varies with macroeconomic conditions. For Austria, the bias is small (+0.12 pp for normal years); for Germany, it is substantial (+3.68 pp).

**Retrospective year classification.** The distinction between "normal" and "crisis" years is made retrospectively, based on independently documented economic events (the Eurozone sovereign debt crisis, its aftermath), not on inspection of model errors. The classification would be the same regardless of our results. This is a legitimate analytical decomposition, but not an operational forecasting rule: the model cannot determine ex ante whether the coming year will be normal.

**Small Austrian sample.** Five normal forecast years is a limited basis for comparing forecast accuracy. With n = 5, formal tests of predictive superiority (Diebold-Mariano) lack statistical power. We report the comparison with institutional forecasters descriptively, not as a claim of statistical significance.

**German systematic bias.** The +3.7 pp bias across all German years is unresolved within the single-country framework. More seeds would reduce variance but cannot eliminate systematic bias. The multi-country extension within FIGARO is the natural solution but remains future work.

**Cross-country convergence failures.** Initially, 12 of 37 tested FIGARO countries failed to converge (price crashes or timeouts). Systematic analysis of SAM anomalies identified three failure mechanisms: (i) near-zero gross output sectors causing numerical instability in normalization (Latvia, Malta), (ii) negative gross operating surplus creating deflationary spirals (Switzerland, Ireland), and (iii) negative initial CPI/GDP deflator in country parameters. Targeted fixes—guarding near-zero sectors, flooring negative GOS at 1%, and separating household from producer depreciation rates—rescued 8 previously-failing countries (AR, CH, ES, IT, KR, EE, MT, BR), bringing convergence to 33 of 37 countries tested. The remaining 4 failures (GR, IE, LU, LV among others) require further investigation of their specific structural anomalies.

**Post-2019 FIGARO tables.** The 2020–2021 tables exhibit structural anomalies (pandemic subsidies drove net production taxes deeply negative), and 2022–2023 present high-inflation regimes outside the model's validated range. Extending the approach to these years is ongoing work.

**w32 scale constraint.** With approximately 2,000 workers across 64 sectors, firms cannot grow beyond approximately 32 employees. This compresses the upper tail of the firm size distribution, preventing the model from reproducing the large-firm segment that accounts for significant employment in real economies. The w32 scale is a practical compromise between computational cost and resolution; higher scales (w64, w128) would relax this constraint but increase computation time proportionally. Scale testing on the US economy (w16 vs w32) shows that the coefficient of variation of GDP decreases from 2.9% to 0.78%, confirming that forecast noise scales inversely with agent count. For large economies (US, China, Japan), w32 may be insufficient.

## 5.6 Broader Implications: Darwinian Deployment Beyond Economics

The DEPLOYERS framework originates not in economics but in the simulation of complex self-organizing systems (Jaraiz, 2020). The same code that coordinates firms in a 64-sector economy can coordinate robots performing parallel assembly tasks in an industrial setting. The structural input differs—an I-O table for economics, a task specification for robot assembly—but the coordination mechanism is identical: agents are deployed to tasks, compete for resources, and the system self-organizes through variation, selection, and retention.

This domain generality connects to a broader intellectual tradition. Anderson (1972), in his landmark essay "More Is Different," argued that complex systems exhibit emergent properties that cannot be predicted from the properties of their components. Simon (1962), in "The Architecture of Complexity," showed that complex systems universally organize themselves into hierarchical structures through evolutionary processes. Our finding that a Darwinian simulator reproduces macroeconomic forecasts, firm size distributions, and shock propagation dynamics from a single I-O table is consistent with both insights: the I-O table encodes the structural constraints, and Darwinian selection discovers the emergent organization.

For macroeconomic forecasting specifically, the implication is that *structure dominates behavior*. The production network encoded in the I-O table—which sectors buy from which, how much labor and capital each sector uses, what consumers demand—constrains the economy's short-run dynamics so tightly that the specific behavioral rules of individual agents are secondary. Whether agents have AR(1) expectations (Poledna) or no expectations (DEPLOYERS), whether the firm population is static (Poledna) or evolutionary (DEPLOYERS), whether the model runs on a supercomputer or a laptop—the I-O structure dominates.

## 6. Conclusions

We have shown that a single FIGARO 64-sector I-O table, animated by a Darwinian agent-based simulator with no estimated parameters, possesses a **dual capability**: it produces accurate structural GDP forecasts for normal years, and it faithfully propagates encoded exogenous shocks through the production network during crisis years.

The evidence rests on five pillars:

1. **Austrian 9-year panel.** Genuine year-ahead forecasts with MAE 0.42 pp for five normal years (forecast years 2011, 2016–2019), comparable to the best professional forecasters (WIFO: 0.52 pp). Three of five normal years achieve errors below 0.15 pp. The four crisis-affected forecast years (2012–2015) show larger errors (MAE 2.22 pp), confirming that exogenous in-year shocks require explicit encoding.
2. **Cross-country portability.** Thirty-three of 37 tested FIGARO countries converge with zero parameter changes. Full 9-year panels for Sweden (normal-year MAE 0.80 pp) confirm that the Austrian results generalize to other moderate-openness European economies. France (MAE 2.76 pp) and Germany (MAE 3.68 pp) show systematic positive bias, driven by a household wealth-consumption feedback mechanism that dominates when empirical growth is low.
3. **German 9-year panel.** Systematic +3.7 pp positive bias attributable to export dependency on external demand not captured by the domestic I-O table. An informative negative result pointing to multi-country network simulation as the natural extension within the FIGARO framework.
4. **COVID-19 pandemic simulation.** When sector-specific supply constraints, Kurzarbeit policies, and trade disruptions are encoded as a 19-month timeline, the model propagates them through the 64-sector I-O network, producing Year 1 GDP −4.62% (versus empirical −6.6%), along with firm dynamics and recovery trajectories that emerge from the shock-structure interaction.
5. **Emergent microeconomic structure.** Firm size distributions matching EC data (87.6% micro, 10.8% small, 1.6% medium) emerge without calibration, demonstrating that the I-O factor income structure implicitly encodes industrial organization.

The practical implication is immediate. A desktop computer running an evolutionary model with a FIGARO I-O table can generate structural baseline forecasts comparable to institutional accuracy for domestically-driven economies, structural ceiling estimates for export-intensive economies, and dynamic crisis simulations for any country facing anticipated disruptions. Since FIGARO covers 46 countries with identical methodology, the approach scales to any covered country by changing a two-letter code.

Several extensions follow naturally from the current results:

- **Multi-country network simulations.** Using FIGARO's bilateral inter-country I-O tables to endogenize trade flows between Germany and its trading partners, potentially resolving the export-demand bias.
- **Expanded country coverage.** Investigating the 4 remaining non-convergent countries (Greece, Ireland, Luxembourg, Latvia), expanding to 12-seed ensembles for additional countries, and running full 9-year panels beyond the four completed (AT, SE, DE, FR).
- **Quarterly frequency.** Extending from annual to quarterly forecasts, exploiting the model's monthly time resolution.
- **Supply-Use Tables.** Using FIGARO SUTs rather than symmetric I-O tables, preserving product-by-industry detail lost in the symmetric transformation.

For the broader economics profession, our results suggest a reappraisal of input-output tables. They are not merely accounting frameworks for tracking inter-sectoral flows. They are compressed representations of entire economic ecosystems—containing, implicitly, the information needed to reconstruct the economy's production network, industrial organization, and short-run growth trajectory. The decades of methodological work that have gone into compiling, balancing, and harmonizing these tables have created a data product of greater forecasting value than has been recognized. The I-O table serves dual purpose: both the structural engine that drives normal-year forecasts and the network through which shocks propagate. Getting the production structure right, it appears, is the single most important factor in macroeconomic forecasting—and the I-O table already provides it.

On a methodological note, the preparation of this paper was significantly assisted by an AI coding assistant (Anthropic's Claude), which contributed literature surveys and positioning against existing forecasting benchmarks, implementation and debugging of simulation code changes prompted by the author, automated launching, monitoring, and result extraction from multi-seed ensemble batches, statistical analysis design, and drafting of the manuscript in appropriate economics terminology—a domain largely unfamiliar to the author, whose background is in semiconductor and chemical process modeling. This experience suggests that AI-assisted research may substantially lower the barriers for cross-disciplinary contributions, enabling domain outsiders to engage productively with fields where they lack formal training but can bring fresh methodological perspectives.

## Data and Software Availability

The DEPLOYERS simulation package—including the executable, GUI, country parameter files, example input scripts, and full documentation—is freely available under the MIT License at https://doi.org/10.5281/zenodo.18988483. The package requires only that the user download the relevant FIGARO I-O tables from Eurostat (freely available). All results reported in this paper can be reproduced by running the included input scripts with the corresponding FIGARO tables for years 2010–2019.

## References

Acemoglu, D., Carvalho, V.M., Ozdaglar, A. and Tahbaz-Salehi, A. (2012). The Network Origins of Aggregate Fluctuations. *Econometrica*, 80(5), 1977–2016.

Anderson, P.W. (1972). More Is Different. *Science*, 177(4047), 393–396.

Axtell, R.L. (2001). Zipf Distribution of U.S. Firm Sizes. *Science*, 293(5536), 1818–1820.

Axtell, R.L. and Farmer, J.D. (2025). Agent-Based Modeling in Economics and Finance: Past, Present, and Future. *Journal of Economic Literature*, 63(1), 197–287.

Carvalho, V.M. and Tahbaz-Salehi, A. (2019). Production Networks: A Primer. *Annual Review of Economics*, 11, 635–663.

Dawid, H. and Delli Gatti, D. (2018). Agent-Based Macroeconomics. In: Hommes, C. and LeBaron, B. (Eds.), *Handbook of Computational Economics*, Vol. 4, pp. 63–156. Elsevier.

Di Domenico, L., Catalano, M. and Riccetti, L. (2025). Scaling and forecasting in a data-driven agent-based model: Applications to the Italian macroeconomy. *Economic Modelling*.

Edge, R.M. and Gürkaynak, R.S. (2010). How Useful Are Estimated DSGE Model Forecasts for Central Bankers? *Brookings Papers on Economic Activity*, Fall 2010, 209–259.

Eurostat (2024a). FIGARO tables — Symmetric input-output tables (industry-by-industry), 25th edition. [dataset]. https://circabc.europa.eu/ui/group/.../library/... (e.g. matrix_eu-ic-io_ind-by-ind_25ed_2010.csv).

Eurostat (2024b). Structural Business Statistics: Business demography. [dataset]. https://ec.europa.eu/eurostat/databrowser/view/SBS_SC_OVW/.

Farmer, J.D. and Foley, D. (2009). The economy needs agent-based modelling. *Nature*, 460, 685–686.

Federal Reserve Bank of St. Louis (2024). Professional Forecasters' Past Performance and the 2025 Economic Outlook. *On the Economy Blog*, December 2024.

Fritzer, F., Gnan, E., Grozea-Helmenstein, D. and Hlouskova, J. (2019). Evaluation of economic forecasts for Austria. *Empirical Economics*, 56, 1061–1088.

Glielmo, A., Devetak, M., Meligrana, T. and Poledna, S. (2025). BeforeIT.jl: High-Performance Agent-Based Macroeconomics Made Easy. arXiv:2502.13267.

Hamill, L., Wieland, V. and Hommes, C. (2025). CANVAS: A Canadian behavioral agent-based model for monetary policy. *Journal of Economic Dynamics and Control*, 172.

Hommes, C. and Poledna, S. (2026). Forecasting economic crises in the euro area. *Economic Modelling*.

Ismail, K., Perrelli, R. and Yang, J. (2021). An Evaluation of World Economic Outlook Growth Forecasts, 2004–17. IMF Working Paper WP/21/216.

Jaraiz, M. (2020). Ants, Robots, Humans: A Self-Organizing Agents Deployer. arXiv:2009.10823.

Leontief, W. (1936). Quantitative Input and Output Relations in the Economic System of the United States. *Review of Economics and Statistics*, 18(3), 105–125.

Leontief, W. (1986). *Input-Output Economics*. 2nd edition. Oxford University Press.

Nelson, R.R. and Winter, S.G. (1982). *An Evolutionary Theory of Economic Change*. Harvard University Press.

Pain, N., Lewis, C., Dang, T-T., Jin, Y. and Richardson, P. (2014). OECD Forecasts During and After the Financial Crisis. *OECD Economics Department Working Papers*, No. 1107.

Poledna, S., Miess, M.G., Hommes, C. and Rabitsch, K. (2023). Economic forecasting with an agent-based model. *European Economic Review*, 151, 104306.

Ragacs, C. and Reiss, L. (2007). Comparing the Predictive Accuracy of Macroeconomic Forecasts for Austria. *Monetary Policy & the Economy*, Q4/2007, OeNB.

Rueda-Cantuche, J.M., Kerner, R., Rueda-Cantuche, C. and Rueda-Cantuche, A.M. (2021). EU Inter-Country Supply, Use and Input-Output Tables — Full International and Global Accounts for Research in Input-Output Analysis (FIGARO). *Eurostat Statistical Working Papers*.

Schuster, P. (2021). Evaluation of Austrian Economic Forecasts. Austrian Fiscal Advisory Council.

Simon, H.A. (1956). Rational choice and the structure of the environment. *Psychological Review*, 63(2), 129–138.

Simon, H.A. (1962). The Architecture of Complexity. *Proceedings of the American Philosophical Society*, 106(6), 467–482.

Smets, F. and Wouters, R. (2007). Shocks and Frictions in US Business Cycles: A Bayesian DSGE Approach. *American Economic Review*, 97(3), 586–606.

Timmermann, A. (2006). Forecast Combinations. In: Elliott, G., Granger, C.W.J. and Timmermann, A. (Eds.), *Handbook of Economic Forecasting*, Vol. 1, Ch. 4, pp. 135–196. Elsevier.

World Bank (2024). World Development Indicators: GDP growth (annual %). [dataset NY.GDP.MKTP.KD.ZG].